\documentclass[a4paper,twocolumn,showpacs,nofootinbib,aps,showkeys,floatfix]{revtex4}
\input{epsf}
\usepackage{graphicx}
\usepackage{epsfig}

\begin{document}

\title{A parametric model for dark energy}
\author{E. M. Barboza Jr.$^{1}$\footnote{E-mail: edesio@on.br}}

\author{J. S. Alcaniz$^{1,2}$\footnote{E-mail: alcaniz@on.br}}

\address{$^{1}$Observat\'orio Nacional, 20921-400, Rio de Janeiro - RJ, Brasil}

\affiliation{$^{2}$Instituto Nacional de Pesquisas Espaciais/CRN, 59076-740, Natal -- RN, Brasil}

\date{\today}

\begin{abstract}
Determining the mechanism behind the current cosmic acceleration constitutes a major question nowadays in theoretical physics. If the dark energy route is taken, this problem may potentially bring to light new insights not only in Cosmology but also in high energy physics theories. Following this approach, we explore in this paper some cosmological consequences of a new time-dependent parameterization for the dark energy equation of state (EoS), which is a well behaved function of the redshift $z$ over the entire cosmological evolution, i.e.,  $z\in [-1,\infty)$. This parameterization allows us to divide the parametric plane $(w_0,w_1)$ in defined regions associated to distinct classes of dark energy models that can be confirmed or excluded from a confrontation with current observational data. A statistical analysis involving the most recent observations from type Ia supernovae,  baryon acoustic oscillation peak, Cosmic Microwave Background shift parameter and Hubble evolution $H(z)$ is performed to check the observational viability of the EoS parameterization here proposed.
\end{abstract}


\maketitle

\section{Introduction}

Over the last decade, a considerable number of high quality observational data have transformed radically the field of cosmology. Results from distance measurements of type Ia supernovae (SNe Ia)  \cite{Riess, Astier} combined with Cosmic Microwave Background (CMB) observations \cite{wmap} and the Large-Scale Structure (LSS) data \cite{lss, Eisenstein} seem to indicate that the simple picture provided by the standard cold dark matter scenario is not enough. These observations are usually explained by introducing a new hypothetical energy component with negative pressure, the so-called dark energy or \emph{quintessence}, usually characterized by the equation of state (EoS) parameter $w \equiv p/\rho$, i.e., the ratio between the dark energy pressure to its energy density (see, e.g., \cite{rev} for some recent reviews).

Among the many candidates for dark energy, the energy density associated with the quantum vacuum or the cosmological constant ($\Lambda$) emerges as the  simplest and the most natural possibility. However, this interpretation of the cosmological term brings to light  an unsettled situation in the Particle Physics/Cosmology interface, in which the cosmological upper bound ($\rho_{\Lambda} \lesssim 10^{-47}$ GeV$^4$) differs from theoretical expectations ($\rho_{\Lambda} \lesssim 10^{71}$ GeV$^4$) by more than 100 orders of magnitude \cite{Weinberg}. Thus, although $\Lambda$ may be able to explain the majority of observations available so far, if dark energy is in fact associated with the vacuum energy density, we should look for an explanation for this enormous discrepancy between theory and observation. 

In this regard, many proposals have appeared in the literature trying to solve this problem, but so far no reasonable explanation, if there is one, was obtained. Thus, despite the beauty and simplicity of the cosmological term, other proposals, even if not so attractive, should be explored. In the context of the General Theory of Relativity, for instance, models with a time-varying cosmological term \cite{atcdm}, irreversible processes (e.g., cosmological matter creation \cite{mc}), barotropic fluids (e.g., Chapligyn Gas \cite{Chapligyn}) and, dynamical scalar fields $\Phi$ (e.g., quintessence \cite{Quintessence}, phantom fields \cite{Phantom} and quintom \cite{Quintom}), are some of those alternatives\footnote{Out of the context of the General Relativity, some other attractive approaches to the dark energy problem, such as brane-world models \cite{brane} and f(R) derived cosmologies \cite{fr} have also been recently explored.}. 

Phenomenologically, it is usual to explore some possible time-dependent parameterizations to describe the dark energy EoS.  In this concern, Taylor series-like parameterizations 
%\begin{equation}
$$
\label{TaylorExpantion}
w(z)=\sum_{n=0}w_n\,x_n(z),
%\end{equation}
$$
where $w_n$'s are constants to be fixed by observations and $x_n(z)$'s are functions of redshift, are among the most commonly discussed and, depending on the allowed values for $w_n$'s, they may have quintessence ($-1\leq w(z)\leq1$) and phantom fields ($w(z)<-1$) as special cases. Among the parameterizations based on series expansion we can quote the following first order expansions: 
\begin{eqnarray}
\label{OtherParameterizations}
w(z) = \; \left\{
\begin{tabular}{l}
$w_0 + w_1 z$ %\quad    (P1)  
\, \quad \quad  \quad \quad \mbox{(redshift)} \quad\,  \hspace{0.2cm}\cite{Huterer}\\
\\
$w_0+w_1z/(1+z)$ \quad \mbox{(scale factor)} \, 
\cite{Chevallier} \\
\\
$w_0+w_1\ln(1+z)$ %\quad (P3)  
\quad  \mbox{(logarithmic)} \quad \cite{Efstathiou}
\end{tabular}
\right.
\nonumber
\end{eqnarray}

The first parameterization represents a good fit for low redshifts, but presents a problematic behavior for high redshifts. For example, it fails to explain the estimated ages of high-$z$ objects \cite{Friaca}. The second one solves this problem, since $w (z)$ is a well behaved function for $z\gg1$ and recovers the linear behavior in $z$ at low redshifts. The latter was built empirically to adjust some quintessence models at $z\lesssim4$. 
It is worth mentioning that it is difficult to obtain the above parameterizations from scalar field dynamics since they are not limited functions, i.e., the EoS parameter does not lie in the interval defined by $w = \frac{{\dot{\Phi}^2/2} - V(\Phi)}{{\dot{\Phi}^2}/{2} + V(\Phi)}$, where $V(\Phi)$ is the field potential.
This amounts to saying that when extended to the entire history of the universe, $z\in [-1,\infty)$, the three parameterizations above are divergent functions of the redshift\footnote{As an example, the scale factor parameterization above blows up exponentially in the future as $z \rightarrow -1$ for $w_1 > 0$  --- see, e.g., \cite{teg} for a discussion.}. However, since the dark energy dominance is a very recent phenomena, this particular aspect is not usually taken as important because it is always possible to obtain a quintessence-like behavior as a particular approximation when $z$ is not too large. Even so, one can suspect that the information that can be obtained about dark energy from these parameterizations may be compromised. In fact, to avoid the ambiguities and uncertainty that can be contained in these parameterizations, it is desirable a parameterization that can be extended to entire expansion history of the Universe, so that the constraints obtained from the scalar field behavior can also be applied.

In this paper, in order extend the range of  applicability of the dark energy EoS, we study some cosmological consequences of a new phenomenological parameterization for this quantity. We discuss the classification of this expression in the parametric space $w_0 - w_1$, and explore its main observational features. We test the viability of this new dark energy scenario from the most recent distance measurements from type Ia supernovae (SNe Ia), measurements of the  baryonic acoustic oscillations (BAO), the shift parameter of the cosmic microwave background and measurements of the rate of the cosmic expansion $H(z)$.

\section{Parameterization}

In this paper, we will consider the following parameterization for the dark energy EoS:
\begin{equation}
\label{MyParameterization}
w(z)=w_0+w_1\frac{z(1+z)}{1+z^2},
\end{equation}
where $w_0$ is the EoS value at present time (the subscript and superscript zero denotes the present value of a quantity) and $w_1=dw/dz\vert_{z=0}$ gives a measure of how time-dependent is the dark energy EoS. This parameterization has the same linear behavior in $z$ at low redshifts presented by the parameterizations discussed above and has the advantage of being a limited function of $z$ throughout the entire history of the Universe. 

For a flat Friedmann-Robertson-Walker universe, it is straightforward to show from the continuity equation for each component, $\dot{\rho}_i=3\,\rho_i\,\dot{z}[1+w_i(z)]/(1+z)$, that the dark energy density $\rho_{X}$ for parameterization (\ref{MyParameterization}) evolves as
\begin{equation}
\label{DE-Density}
f(z)\equiv\frac{\rho_{X}}{\rho_{X}^{\,\,0}}=(1+z)^{3(1+w_0)}(1+z^2)^{3w_1/2}\;.
\end{equation}
Note that, if Eq. (\ref{MyParameterization}) should be always valid, the parameters $w_0$ and $w_1$ should be constrained by
\begin{equation}
\label{EarlyTimeConstraint}
w_0+w_1<0\;,
\end{equation}
so that the dark energy is always subdominant at $z\gg1$. 

In this background, the Friedman equation for a dark matter/dark energy dominated universe reads
\begin{equation}
\label{FriedmannEquation}
H^2\equiv\Big(\frac{\dot{a}}{a}\Big)^2=H_0^2[\Omega_m^0(1+z)^3+(1-\Omega_m^0)\,f(z)]\;,
\end{equation}
where $a$ is the cosmological scale factor, $H$ is the Hubble parameter and $\Omega_i^0=\rho_i^0/\rho_c^0$ ($\rho_c^0=3\,H_0^2/8\,\pi\,G$) is density parameter of the $i$th-component ($m \equiv$ baryonic + dark  matter).

The deceleration parameter is given by
\begin{equation}
\label{DeccelerationParameter}
q(z)=-\frac{\ddot{a}}{a\,H^2}=\frac{1}{2}[1+3\,w(z)\,\Omega_X]\;.
\end{equation}
By solving the above equation for $w(z)$ at $z=0$ we obtain
\begin{equation}
\label{GeneralConstraint}
w_0=\frac{2\,q_0-1}{3(1-\Omega_m^{\,0})} \quad \mbox{or} \quad  w_0 <-\frac{1}{3(1-\Omega_m^{\,0})}\;,
\end{equation}
since $q_0<0$, as indicated by current observations \cite{Riess}. Note that, for $\Omega_m^{\,0}=0.27\pm 0.04$ \cite{wmap}, it is possible obtain a model-independent bound on the current value of the dark energy EoS, i.e., $w_0<-0.43$.

\subsection{The $w_0 - w_1$ plane}

By differentiating Eq. (\ref{MyParameterization}) with respect to $z$, we find that $w(z)$ has absolute extremes at $z_{\pm}=1\pm\sqrt{2}$ corresponding, respectively, to $w_-=w(z_-)=w_0-0.21w_1$ and $w_+=w(z_+)=w_0+1.21w_1$. For $w_1>0$ $(<0)$, $w_-$ is a minimum (maximum) and $w_+$ is a maximum (minimum). Since quintessence and phantom scalar fields EoS are limited by $-1\leq w(z)\leq1$ and $w(z)<-1$, respectively, the region occupied in the $(w_0,w_1)$ plane by these fields can be easily determined  by imposing that the maximum and minimum of $w(z)$ satisfy these limits. Thus, for quintessence fields we find the following bounds
$$
-1\leq w_0-0.21w_1\quad  {\mbox{and}} \quad  w_0+1.21w_1\leq1 \quad ({\mbox{if}} \quad w_1>0) \;,
$$
and
$$
-1\leq w_0+1.21w_1\quad  {\mbox{and}} \quad  w_0-0.21w_1\leq1 \quad ({\mbox{if}} \quad w_1<0)\;.
$$
whereas for phantom  fields we obtain 
$$
w_1<-(1+w_0)/1.21 \quad ({\mbox{if}} \quad w_1>0)\;,
$$
and
$$
w_1>(1+w_0)/0.21  \quad ({\mbox{if}} \quad w_1<0)\;.
$$

Figure 1 shows different classes of dark energy models in the $(w_0,w_1)$ plane  that arises from parameterization (\ref{MyParameterization}). The forbidden region represents the constraint (\ref{EarlyTimeConstraint}) while the decelerating region is limited by the upper bound (\ref{GeneralConstraint}) with $\Omega_m^{\,0}=0.27$, i.e., $w_0<-0.43$. The blank regions indicate models that at some point of the cosmic evolution, $z\in [-1,\infty)$, have switched or will switch from quintessence to phantom behaviors or vice-versa. Clearly, Parameterization (\ref{MyParameterization}) provides a simple way to classify distinct DE models in the $(w_0,w_1)$ plane. This is a direct consequence from the fact that (\ref{MyParameterization}) is a well behaved bounded function along the entire history of the Universe. The standard $\Lambda$CDM scenario corresponds to the intersection point between quintessence ($w > -1$) and phantom models ($w < -1$).

\begin{figure}
\centerline{\psfig{figure=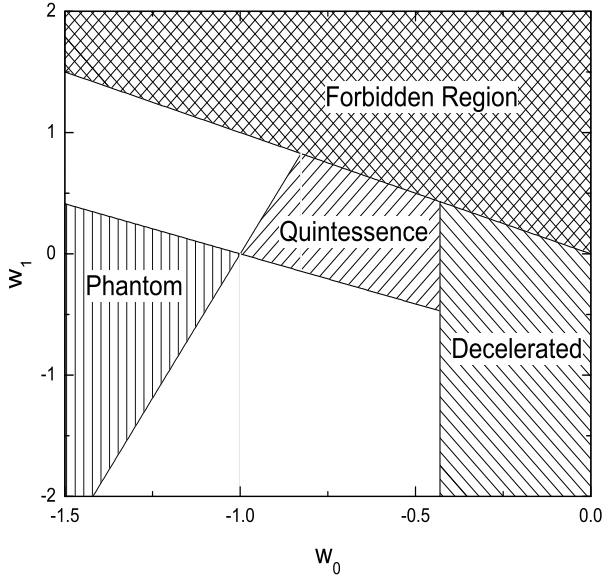,width=3.4truein,height=3.4truein}
\hskip 0.1in} 
\caption{ The $(w_0,w_1)$ parametric space for Parameterization (\ref{MyParameterization}). The forbidden region represents the constraint (\ref{EarlyTimeConstraint}) while the decelerating region is limited by the upper bound (\ref{GeneralConstraint}) with $\Omega_m^{\,0}=0.27$, i.e., $w_0<-0.43$. The blanc regions indicate models that at some point of the cosmic evolution, $z\in [-1,\infty)$, have switched or will switch from quintessence to phantom behaviors or vice-versa.}
\label{w_0-w_1Plane}
\end{figure}

\section{Observational constraints}

In the previous Section, we have defined the regions occupied by different classes of dark energy models derived from parameterization (\ref{MyParameterization}) in the plane $(w_0,w_1)$. Now, we will test the viability of these scenarios by using the most recent cosmological data, namely, $115$ distance measurements of SNe Ia from the Supernova Legacy Survey (SNLS) \cite{Astier}, measurements of the  baryonic acoustic oscillations (BAO)  from SDSS \cite{Eisenstein}, the CMB shift parameter as given by the WMAP team \cite{wmap} and estimates of the Hubble parameter $H(z)$ obtained from ages of high-$z$ galaxies \cite{Simon} (for more  details on the statistical analysis discussed below we refer the reader to Ref.~\cite{refs}).

\subsection{SNe Ia observations}

The predicted distance modulus for a supernova at redshift $z$, given a set of
parameters $\mathbf{P}$, is
\begin{equation} \label{dm}
\mu_p(z|\mathbf{P}) = m - M = 5\,\mbox{log} d_L + 25,
\end{equation}
where $m$ and $M$ are, respectively, the apparent and absolute magnitudes, and $d_L$ stands for the luminosity distance (in units of megaparsecs),
\begin{equation}
\label{LuminosityDistance}
d_L(z;{\bf P})=(1+z)\int_{0}^{z}\frac{dz^{\prime}}{H(z^{\prime};{\bf P})}\;,
\end{equation}
where $H(z; {\bf P})$ is given by Eq. (\ref{FriedmannEquation}).

We estimated the best fit to the set of parameters $\mathbf{s}$ by using a $\chi^{2}$ statistics, with
\begin{equation}
\chi^{2}_{SNe} = \sum_{i=1}^{N}{\frac{\left[\mu_p^{i}(z|\mathbf{P}) -
\mu_o^{i}(z)\right]^{2}}{\sigma_i^{2}}},
\end{equation}
where $\mu_p^{i}(z|\mathbf{P})$ is given by Eq. (\ref{dm}), $\mu_o^{i}(z)$ is the extinction corrected distance modulus for a given SNe Ia at $z_i$, and $\sigma_i$ is the uncertainty in the individual distance moduli. Since we use in our analyses the SNLS collaboration sample (see \cite{Astier} for details),  $N = 115$. 

\begin{figure*}
\centerline{\psfig{figure=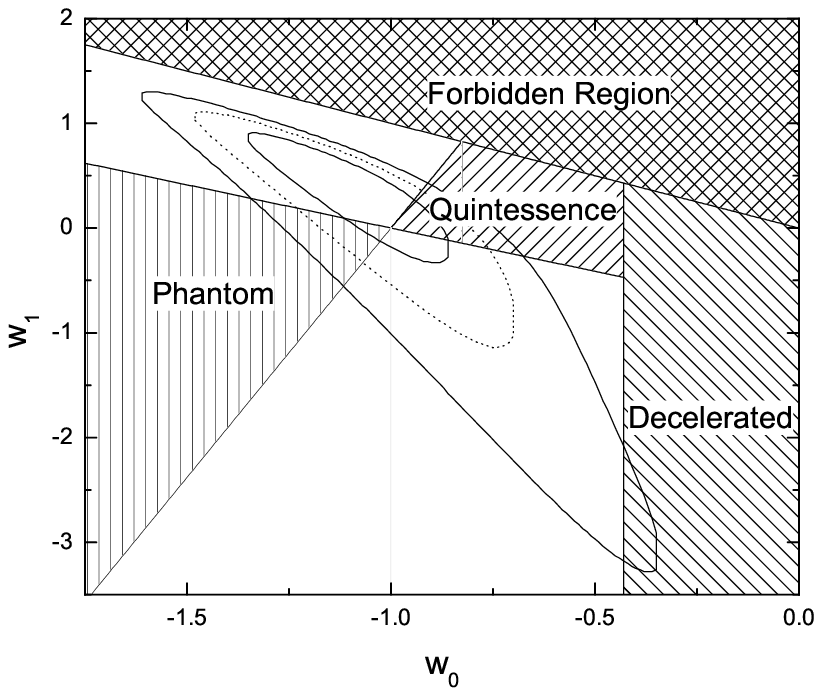,width=3.4truein,height=3.4truein}
\hskip 0.2in
\psfig{figure=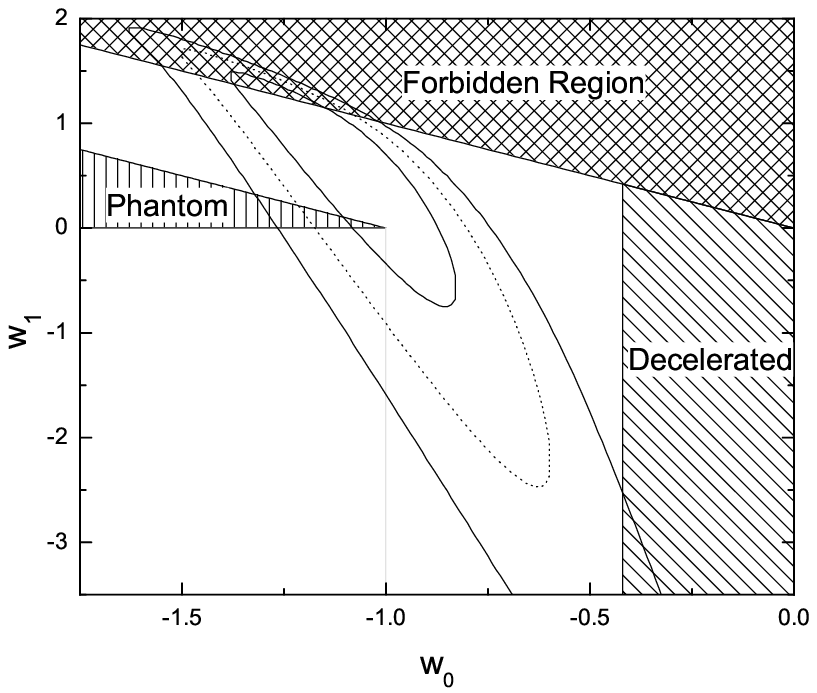,width=3.4truein,height=3.4truein}
\hskip 0.1in} \caption{The results of our statistical analyses. {\bf{Left:}} The 68.3\%, 95.4\%, and 99.7\% confidence contours for Parameterization (1) arising from SNLS SNe Ia, SDSS BAO,  WMAP CMB shift parameter and $H(z)$ data.  {\bf{Right:}} The same as in the previous Panel for the scale factor parameterization.}
\label{fig:zt}
\end{figure*}

\subsection{Baryonic acoustic oscillations}

As well known,  the acoustic peaks in the cosmic microwave background (CMB) anisotropy power spectrum is an efficient way for determining cosmological parameters (e.g., \cite{wmap}).  Because the acoustic oscillations in the relativistic plasma of the early universe will also be imprinted on to the late-time power spectrum of the non-relativistic matter \cite{peeblesyu}, the acoustic signatures in the large-scale clustering of galaxies yield additional tests for cosmology. 

In particular, the characteristic and reasonably sharp length scale measured at a wide range of redshifts provides an estimate of the distance-redshift relation, which is a geometric complement to the usual luminosity-distance from SNe Ia. Using  a large spectroscopic sample of 46,748 luminous, red galaxies covering 3816 square degrees out to a  redshift of $z=0.47$ from the Sloan Digital Sky Survey, Eisenstein et al.  \cite{Eisenstein} have successfully found the peaks, described by the ${\cal{A}}$-parameter, i.e., 
\begin{equation}
\label{BAO}
\mathcal{A}\equiv\sqrt{\Omega_m^0}\Big[\frac{H_0^2}{z_{\ast}\,H^{2}(z_{\ast};{\bf P})}\int_0^{z_{\ast}}\frac{H_0\,dz}{H(z;{\bf P})}\Big]^{2/3},
\end{equation}
where $z_{\ast}=0.35$ is the redshift at which the acoustic scale has been measured.

\subsection{CMB shift parameter}

The shift parameter  $R$ which determines the whole shift of the CMB angular power spectrum is given by \cite{b}
\begin{equation}
\label{ShiftParameter}
\mathcal{R}\equiv\sqrt{\Omega_m^0}\int_0^{z_{\mathrm ls}}\frac{H_0\,dz}{H(z;{\bf P})},
\end{equation}
where the $z_{\mathrm ls} = 1089$ is the redshift of the last  scattering surface, and the current estimated value for this quantity is ${\mathcal R}_{\mathrm{obs}} = 1.70\pm 0.03 $ \cite{wang}.

\subsection{Hubble Expansion}

In our joint analysis we also use 9 determinations of the Hubble parameter as a function of redshift, as given in Ref. \citep{Simon}. These determinations, based on differential age method, relates the Hubble parameter $H(z)$ directly to measurable quantity  $dt/dz$ and can be achieved from the recently released sample of old passive galaxies from Gemini Deep Deep Survey (GDDS) \citep{Abraham} and archival data \citep{Dunlop}. The use of these data to constrain cosmological models is interesting because, differently of luminosity distance measures, the Hubble parameter is not integrated over (see \citep{Simon} for more details). To perform this test we minimize the quantity
\begin{equation}
\chi^{2}_{H} = \sum_{i=1}^{9}{\frac{\left[H^{i}(z|\mathbf{P}) - H_{obs}^{i}(z)\right]^{2}}{\sigma_i^{2}}},
\end{equation}
where the predicted Hubble evolution for parameterization (\ref{MyParameterization}) is given by Eq. (\ref{FriedmannEquation}).

\subsection{Analyses and Discussions}

\subsubsection{Parameterization (1)}

Figure (2) shows the main results of our analyses. In order to compare the theoretical frame with the observational constraints discussed above the three dimensional parameter space $(\Omega_m^0,w_0,w_1)$ has been projected into the plane $(w_0,w_1)$. In Fig. (2a) we show confidence intervals (68.3\%, 95.4\% and 99.7\%) in this parametric space $(w_0,w_1)$ for parameterization (\ref{MyParameterization}). The best-fit values for these parameters are $w_0=-1.11$ and $w_1=0.43$ whereas at 68.3\% (c.l.) they lie, respectively, in the intervals $-1.35 \leq w_0 \leq -0.86$ and $-0.33 \leq w_1 \leq 0.91$. Note that no DE model is preferred or ruled out by observations, although the largest portion of the confidence contours lies into the blanc region (indicating models that  have switched or eventually will switch from quintessence to phantom behaviors or vice-versa). At $99\%$ C.L. we also have $0.21\leq\Omega_m^0\leq0.33$ so that the decelerated region is limited, in accordance with (\ref{GeneralConstraint}), by the constraint $w_0<-0.42$ and the possibility of a decelerated universe today is almost completely excluded. For the best-fit values discussed above, the transition redshift $z_t$, at which the Universe switches from deceleration to acceleration, occurs at $z_t \simeq 0.58$.

\subsubsection{Scale factor Parameterization}

For the sake of comparison, we employ the same analysis to scale factor parameterization $w_0+w_1z/(1+z)$ \cite{Chevallier}. Note that this  parameterization has an absolute extreme in $w_{\infty}=w(z=\infty)=w_0+w_1$. For $w_1>0$, $w_{\infty}$ is a maximum whereas for $w_1<0$ it is a minimum. Thus, the region occupied by phantom fields is determined by the constraints $w_{\infty}<-1$ and $w_1>0$, whereas similar constraints cannot be obtained for the quintessence case. The region occupied by phantom fields in the context of this parameterization and the confidence intervals for the statistical analysis discussed above are displayed in the Fig. 3. At 68.3\% c.l., we found $w_0=-1.14^{+0.31}_{-0.24}$, $w_1=0.84^{+0.65}_{-1.59}$ and $\Omega_m^0=0.27\pm0.03$. As can be seen, at this confidence level, the possibility of an early dark energy dominance is not completely excluded.

Finally, we also note that for this parameterization, the continuity equation solves as 
\begin{equation}
\label{ScaleFactorDE}
f_{SF}(z)=(1+z)^{3(1+w_0+w_1)}e^{-3w_1\,z/(1+z)}
\end{equation}
so that it must also be submitted to the constraint (\ref{EarlyTimeConstraint}). Therefore, when $z\to-1$ ($a\to\infty$), $f_{SF}(z)$ blows up if $w_1>0$ while $f(z)$, given by Eq. (1), blows up if $w_0<-1$. Thus, the roles of the parameters $w_0$ and $w_1$ are inverted in these scenarios, in the sense that while for Parameterization (1) the fate of the Universe is dictated by the equilibrium part ($w_0$), for the scale factor parameterization the future of the Universe is driven by the time-dependent term $w_1$.

\section{Final Remarks}

The recent cosmic expansion history has the potential to greatly extend our physical understanding of the Universe. This in turn is closed related to the origin and nature of the mechanism behind the current cosmic acceleration, for instance, if it is associated with a new component of energy or large-scale modifications of gravity. 

In this paper, by assuming a hypothetical component of dark energy as the fuel that drives the acceleration of the Universe, we have proposed and studied some theoretical and observational aspects of a new parameterization for this dark energy EoS, as given by Eq. (\ref{MyParameterization}). This parameterization is a well-behaved, bounded function of the redshift throughout the entire cosmic evolution, which allows us to study the effects of a time varying EoS component to the distant future of the Universe at $z = -1$ as well as back to the last scattering surface of the CMB. We have classified different classes of dark energy models in the parametric space ($w_0 - w_1$) and studied their theoretical and observational consequences. In order to check the observational viability of the phenomenological scenario proposed here, we also have performed a joint statistical analysis involving some of the most recent cosmological measurements of SNe Ia, BAO peak, CMB shift parameter and the Hubble expansion $H(z)$. From a pure observational perspective, we have shown that both quintessence and phantom behaviours are fully acceptable regimes, although the largest portion of the confidence contours arising from these observations lies in the region of models that have crossed or will eventually cross these regimes at some point of the cosmic evolution.

\begin{acknowledgments}
The authors are grateful to R. Silva for many useful suggestions and a critical reading of the manuscript. E.M.B. Jr. is partially supported by CNPq. JSA is supported by CNPq under Grants 304569/2007-0 and 485662/2006-0.
\end{acknowledgments}

\end{document}